\documentclass[%
 reprint,
 amsmath,amssymb,
 aps,
]{revtex4-1}

\usepackage{upgreek}

\newcommand{\tens}[1]{\tilde{#1}}
\newcommand{\Hext}{\vec H_{\mathrm{ext}}}
\newcommand{\mHext}{H_{\mathrm{ext}}}
\newcommand{\Xm}{\chi_0}
\newcommand{\Msat}{M_\mathrm{sat}}
\newcommand{\Mfk}{\vec M_{i}}

\newcommand{\Hint}{\vec H_{\mathrm{int}}}
\newcommand{\mHint}{H_{\mathrm{int}}}

\usepackage{hyperref}
\newcommand{\HME}{HSME}

\usepackage{graphicx}
\usepackage{subcaption}
\usepackage{dcolumn}

\usepackage{amsmath}
\usepackage{amssymb}

\usepackage[section]{placeins}

\begin{document}

\title{Magnetostriction in elastomers with mixtures of magnetically hard and soft microparticles: effects of non-linear magnetization and matrix rigidity}

\author{Oleg V. Stolbov}%
\affiliation{%
Laboratory of Physics and Mechanics of Soft Matter, Institute of Continuous Media Mechanics, Russian Academy of Sciences (Ural Branch), Perm, Russia
}%

\author{Pedro A. S\'anchez}
\affiliation{%
 Institute of Natural Sciences and Mathematics, Ural Federal University, Ekaterinburg, Russian Federation
}%

\author{Sofia S. Kantorovich}
\affiliation{%
 Institute of Natural Sciences and Mathematics, Ural Federal University, Ekaterinburg, Russian Federation
}%
\altaffiliation[Also at ]{
 Computational and Soft Matter Physics, University of Vienna, Vienna, Austria
}%


\author{Yuriy L. Raikher}
\affiliation{%
Laboratory of Physics and Mechanics of Soft Matter, Institute of Continuous Media Mechanics, Russian Academy of Sciences (Ural Branch), Perm, Russia
}




\begin{abstract}
In this contribution a magnetoactive elastomer (MAE) of mixed content, \textit{i.e.}, a polymer matrix filled with a mixture of magnetically soft and magnetically hard spherical particles, is considered. The object we focus at is an elementary unit of this composite, for which we take a set consisting of a permanent spherical micromagnet surrounded by an elastomer layer filled with magnetically soft microparticles. We present a comparative treatment of this unit from two essentially different viewpoints. The first one is a coarse-grained molecular dynamics simulation model, which presents the composite as a bead-spring assembly and is able to deliver information of all the microstructural changes of the assembly. The second approach is entirely based on the continuum magnetomechanical description of the system, whose direct yield is the macroscopic field-induced response of the MAE to external field, as this model ignores all the microstructural details of the magnetization process. We find that, differing in certain details, both frameworks are coherent in predicting that a unit comprising magnetically soft and hard particles may display a non-trivial re-entrant (prolate/oblate/prolate) axial deformation under variation of the applied field strength.
  The flexibility of the proposed combination of the two complementary frameworks enables us to look deeper into the manifestation of the magnetic response: with respect to the magnetically soft particles, we compare the linear regime of magnetization to that with saturation, which we describe by the Fr\"{o}hlich-Kennelly approximation; with respect to the polymer matrix, we analyze the dependence of the re-rentrant deformation on its rigidity.

\end{abstract}

\pacs{Valid PACS appear here}
\maketitle


\section{Introduction}
\label{sec:intro}
The embedding of solid micro- and nano-particles with magnetic properties into elastic polymer matrices became in recent years one of the most successful approaches for the design of `smart' materials, \textit{i.e.}, materials with a predefined response to external stimuli \cite{2007-filipcsei, 2010-stuart}. The addition of the magnetic component allows to control on-the-fly the rheological properties of the viscoelastic polymer medium by means of applied external fields. Closely related to ferrofluids and magnetorheological fluids \cite{2006-odenbach, 2011-devicente, 2016-odenbach}, these polymer-based materials include gels, whose structure is swollen by a liquid background \cite{1995-shiga, 1997-zrinyi, 2000-zrinyi, 2013-thevenot, 2015-roeder}, and elastomers, that are dry rubber-like materials \cite{1996-jolly, 1999-ginder, 2005-gong-pt, 2006-varga, 2007-fuchs, 2010-chertovich, 2016-odenbach}. The latter, known generically as magnetic elastomers (MEs), are attracting large research efforts in recent years due to the broad range of applications that their magnetically controlled physical properties are inspiring \cite{2000-carlson, 2008-li, 2013-li, 2013-stepanov-jpcs, 2014-li, 2015-ubaidillah, 2018-shamonin}. A main part of such applications are related to their strong response to external fields, that leads to large variations of their shape and mechanical properties. For example, MEs are used to design controllable vibrational absorbers and mounts with tunable stiffness \cite{2006-deng, 2007-abramchuk, 2008-sun}, soft actuators and micromanipulators \cite{2012-boese}, force sensors and artificial muscles for soft robotics \cite{2017-volkova-jmmm, 2019-becker-aam}, coatings with tunable wettability \cite{2015-huang-am, 2018-sorokin} or optical properties \cite{2018-koch}, tunable radiation absorbers \cite{2019-kuznetsova-mtc} or biomedical implants \cite{2019-alekhina-japs}.

The overall response of MEs to external fields is determined by a mechanically constrained but substantial rearrangement of their embedded magnetic particles \cite{2014-gundermann, 2016-odenbach}: for instance, under uniform external fields, magnetically hard particles possessing a permanent magnetic moment tend to align with the direction of the field, whereas magnetically soft particles tend to acquire induced magnetic moments in the same direction; in both cases, if the particle density is large enough to let interparticle interactions to be significant, particles will tend to assemble into straight chains parallel to the field. However, such rearrangements necessarily involve some degree of local deformation of the polymer matrix, either elastic \cite{2014-gundermann, 2017-schuemann-sms} or inelastic \cite{2018-sanchez-sm1}. Therefore, macroscopic changes in the properties of MEs as a response to external fields are the result of the interplay between the field-induced assembly of their magnetic particles and the mechanical constraints of the polymer matrix. Such changes include giant magnetorehological effects entailing large field-induced increases of the elastic modulus \cite{2002-ginder-ijmpb, 2005-gong-pt, 2006-abramchuk, 2010-chertovich, 2014-stoll} and large magnetostriction effects, corresponding to variations in the shape of the sample \cite{2002-ginder-ijmpb, 2006-bednarek, 2008-guan, 2013-stepanov-jpcs}. Under elastic regimes, magnetostriction in MEs is oftent fully reversible, leading to magnetic shape memory effects \cite{2007-boese, 2007-stepanov}.

Even though the measurement of macroscopic magnetorheological and magnetostriction effects with current techniques is straightforward, the accurate fundamental characterization of their underlying microscopic mechanisms and, thus, the rational design of materials with taylored sophisticated properties, still is a serious challenge. Direct observations of the field-induced microstructural changes within MEs became available only in recent years, either by means of optical microscopy \cite{2002-bellan,2008-stepanov,2007-abramchuk,2007-stepanov,2012-an} or by cutting-edge techniques such as X-ray computerized microtomography \cite{2012-guenther, 2017-schuemann-sms, 2018-sanchez-sm1, 2019-watanabe-ijms}. However, so far these techniques provide only static information on the internal microstructure. Application of particle tracking methods \cite{2012-borbath,2014-gundermann, 2018-pessot-jpcm} is a promising experimental approach to achieve dynamic characterization that is still under development.

Following the growing interest in these systems, large theoretical research efforts have been devoted to MEs in recent years. Classical approaches to the modeling of rubber-like materials are based on numerical solving of constitutive equations describing their elastic properties. In the case of MEs, such continuum description can be applied not only to the polymer matrix, but also to the distribution of embedded magnetic particles. This implies to define, on the basis of microscopic motivations or phenomenological approaches, constitutive equations for both, the elastic and the magnetic properties \cite{2003-brigadnov-ijss, 2004-dorfmann-cms}. The simplest approximations within this framework assume a linear elastic behavior, along with linear or nonlinear magnetic properties, and a weak magnetoelastic coupling. The latter implies treating the magnetic forces as mechanical loads, which allows to solve the mechanical and magnetic equations separately. More accurate approaches involve taking into account the nonlinearity of elastic response at finite strains \cite{2003-dorfmann-ejmas, 2004-dorfmann-am,2004-dorfmann-ijnlm,2006-bustamante-qjmam, 2008-raikher-jpcm, 2011-ponte-castaneda-jmps, 2019-mukherjee-ijnlm}. Further important refinements are the consideration of geometry of the boundaries of finite samples \cite{2004-kankanala-jmps, 2007-bustamanet-jem, 2007-hasebe-qjmam, 2008-raikher-jpcm}, anisotropies in the distribution of magnetic particles \cite{2013-zubarev, 2016-kalina-ijss} or strong magnetoelastic couplings that impose simultaneous solving of the elastic and magnetic equations \cite{2007-hasebe-qjmam, 2016-metsch-cms, 2017-kalina, 2018-metsch-aam}. Continuum approaches have the main advantage of enabling direct comparison to macroscopic properties. However, they generally lack detailed descriptions of the material microstructure.

A widely used alternative to full continuum theoretical models is the explicit representation of the magnetic particles based on the dipole approximation \cite{1996-jolly, 2013-han-ijss, 2015-biller-joam, 2017-romeis, 2018-menzel-aam, 2019-khanouki-cpbe}. This allows both, to naturally incorporate microstructural details by means of the discrete distribution of particles and to treat interparticle magnetic interactions as pair potentials, with the main drawback being significantly higher calculation costs. This approach can be combined with different approximations to treat microstructure and interactions, such as bead-spring network representations of the material \cite{2011-ivaneyko, 2015-ivaneyko, 2018-pessot-jpcm, 2019-sanchez-sm1} or hybrid mean field models \cite{2014-ivaneyko, 2016-romeis, 2019-romeis-sm}. The simplest dipolar approach, that assigns a point dipole moment to each magnetic particle, is a reasonable approximation when the density of magnetic particles inside the elastomer is not high and, thus, mutual magnetization between particles is weak. However, different corrections might be needed when such effect is significant \cite{2014-biller, 2015-biller-pre, 2015-biller-joam}.

While the plethora of existing theoretical approaches for the study of MEs keeps growing \cite{2019-mukherjee-ijnlm, 2019-khanouki-cpbe, 2020-nam-pt}, important experimental aspects such as microscopic inelastic responses \cite{2018-sanchez-sm1} or polydispersity of the magnetic component \cite{2019-winger-jmmm} remain poorly studied. In addition, the experimental search for MEs with enhanced or more sophisticated magnetoelastic behaviors is brings in complex characteristics that pose additional theoretical challenges. An example of a ME material of increased complexity is the one obtained by mixing inside the polymer matrix two types of magnetic microparticles, with different sizes and magnetic properties, in order to achieve a combined active and passive magnetic control of the response \cite{2013-borin-joam, 2016-linke, 2019-borin-aam}. The magnetic mixture consists of a relatively low fraction of large spherical microparticles of {NdFeB} alloy, which are magnetically hard (MH), and a high fraction of smaller carbonyl iron microparticles, which are magnetically soft (MS). In such a mixture, both MH and MS particles respond to external fields (active control), whereas MH ones can be permanently magnetized and affect the surrounding MS particles even in the absence of applied field (passive control). Very recently, we introduced the first theoretical study on the behavior of such magnetically hard+soft elastomers (\HME s) \cite{2019-sanchez-sm2}. Using a twofold modeling strategy, that combines a minimal continuum analytical description and a bead-spring computer simulation model, we analyzed the magnetostriction of a representative elementary cell of such material, consisting of a central MH particle surrounded by a cloud of MS ones, being all mechanically interconnected by the elastic matrix. As a first approximation, we assumed linear elasticity and magnetization under weak magnetoelastic coupling conditions for the continuum model, whereas for the bead-spring representation we adopted simple dipolar particles, also with linear magnetization of the MS ones. Both approaches provided the same qualitative behavior for the two cases we analyzed: non-magnetized and magnetized central particle. In the first case, an axial elongation of the elementary volume in the direction of the field, that grows parabolically with the strength of the latter, was established. In the second case, we found an unusual behavior: due to the fact that the field of the central particle breaks the symmetry of the system, a re-entrant axial deformation arises, in which the cell adopts prolate-oblate-prolate shapes as the strength of an applied field antiparallel to the central dipole increases.

On the basis of our preliminary characterization of the magnetostriction of a \HME\ elementary cell \cite{2019-sanchez-sm2}, in this work we analyze several parameters affecting its behavior. Here, we perform such analysis mainly by means of our continuum magnetoelastic description, whereas simulations with the bead-spring model are used only for consistency checks on a single set of parameters. Both models are conveniently modified to study the effects of a nonlinear magnetization of the MS particles. In addition, we also study the effects of a moderate variation in their initial susceptibility and the impact of different rigidities of the elastic matrix. We show that the saturation magnetization of the MS particles only has significant qualitative effects at high rigidities. Moreover, we find that the re-entrant axial deformation tends to be hindered as the elastic modulus is increased, remaining only the axial elongation at high rigidities. Finally, we observe that a moderate decrease of the initial susceptibility tends to favor the re-entrant behavior, broadening the region of deformation into oblate profiles.

\section{System and modeling approaches}
\subsection{System parameters}
Typical \HME\ samples are synthesized using {NdFeB} MH particles of diameter $d_h \approx 50\ \upmu$m and saturation magnetization $M_h \approx 800$~emu. They are combined with MS particles of carbonyl iron with diameter $d_s \approx 5\ \upmu$m and high initial magnetic susceptibility, $\Xm$. Here, we will sample three different values of $\Xm=\lbrace 0.15, 0.2, 0.24\rbrace$, where the highest one corresponds to the limiting value $\Xm^* \sim 3/4\pi$. The volume fraction of MS particles is around $\rho_s \approx 0.3$. In order to study the effects of rigidity of the polymer matrix, here we also sample several values for its shear modulus, $G$, comprised between $10^5$ and $10^7$ dyn/cm$^3$. For the external field, we sample field strengths up to $1.9 \cdot 10^4$~Oe that is the same order of magnitude of typical saturation fields used for these materials \cite{2017-schuemann-sms}. Finally, for our elementary \HME\ cell we take an ideally spherical MH particle and a homogeneous spherical elastic shell around it of $25\ \upmu$m. The latter contains the aforementioned volume fraction of embedded MS particles.

\subsection{Qualitative description}
\begin{figure}[h]
\centering
    \begin{subfigure}[b]{0.45\textwidth}
        \includegraphics[width=4.6cm]{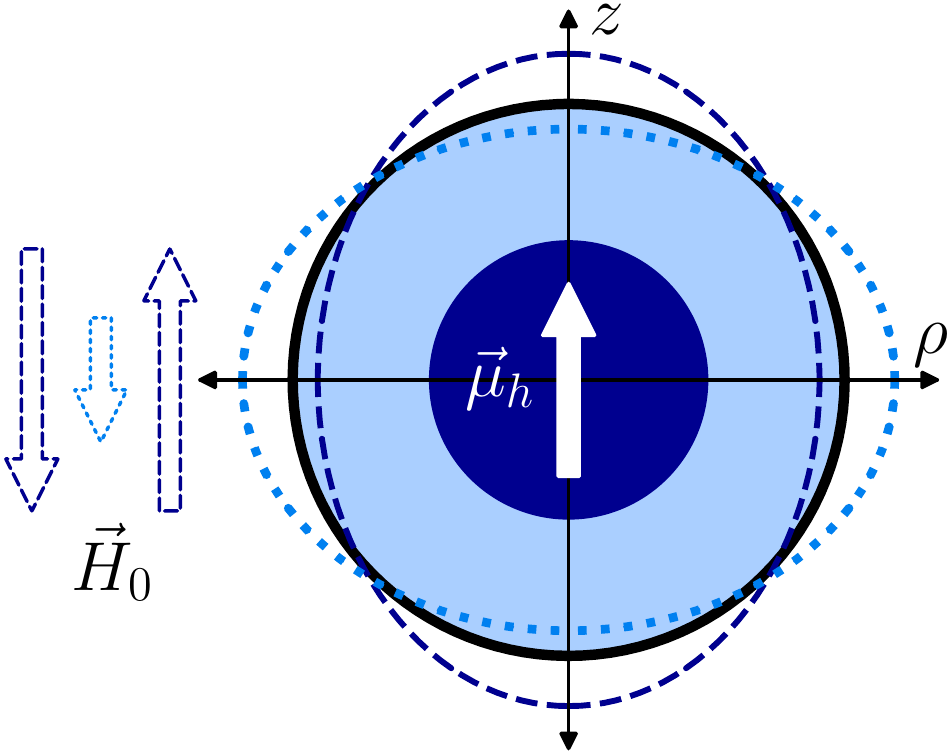}
        \caption{}\label{fig:model-cont}
    \end{subfigure}
    \begin{subfigure}[b]{0.45\textwidth}
        \includegraphics[width=3.5cm]{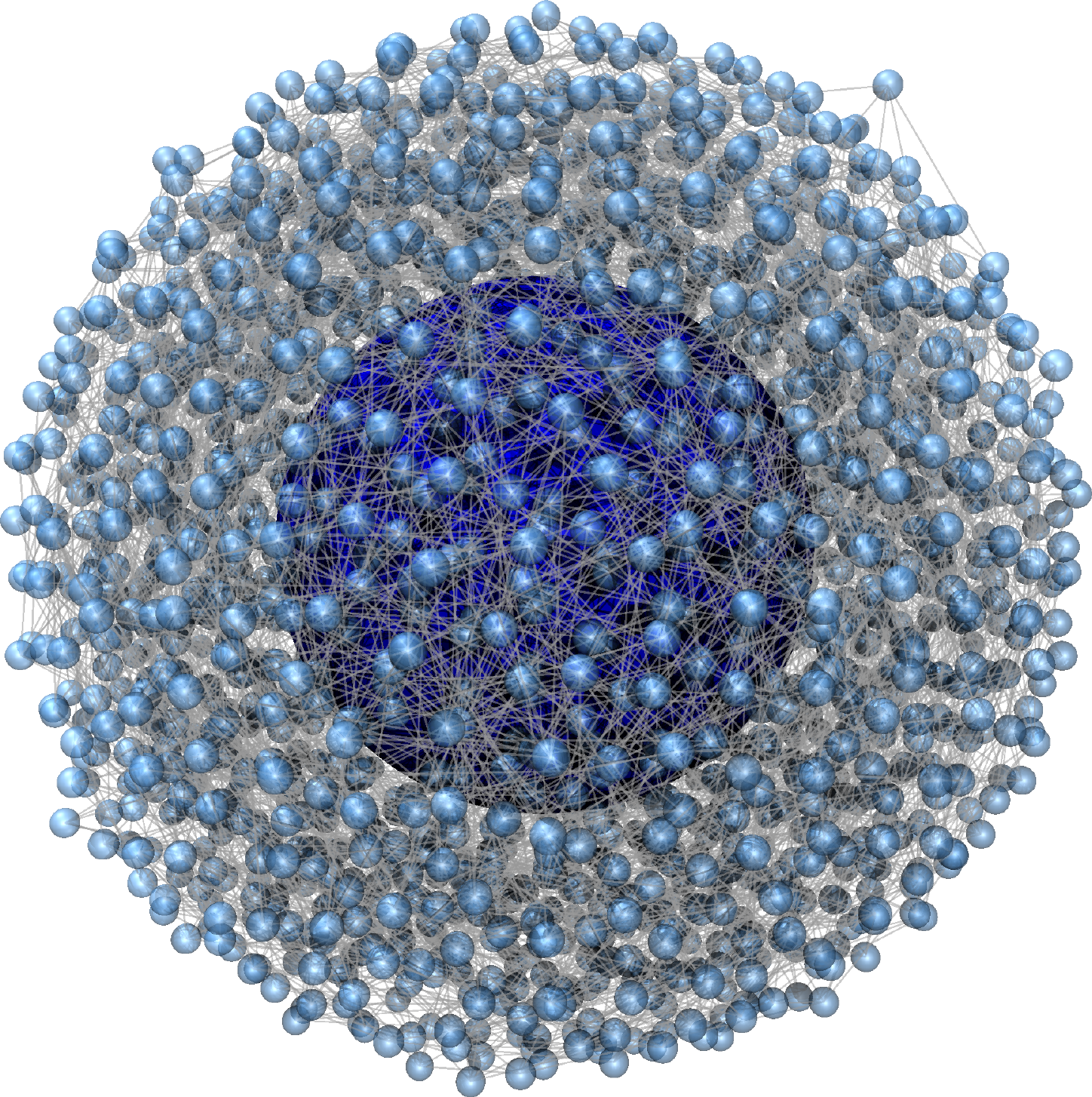}
        \caption{}\label{fig:model-bs}
    \end{subfigure}
\caption{(a) Scheme of the \HME\ elementary cell as represented in the continuum magnetoelastic model. Central dark disc represents the MH particle with dipole moment $\vec \mu_h$ pointing in $z$ direction, light annulus corresponds to the elastic shell with MS particles and its unperturbed boundary is indicated by the solid circle. Dashed and dotted ellipsoids show the shell deformation according to the strength and direction of the external field, $\vec H_0$, indicated by corresponding dashed and dotted arrows. (b) Snapshot of an unperturbed \HME\ elementary cell in the bead-spring simulation model, with a large dark central sphere representing the MH particle and a cloud of small light spheres representing the MS ones, connected by a network of elastic springs depicted as semitransparent lines. Radius of the MS particles has been halved to ease the visualization.}
\label{fig:models}
\end{figure}

Figure~\ref{fig:models} shows the schematic representations of the elementary cell of a \HME\ defined for our analytical magnetoelastic and bead-spring simulation models. Here, we use Figure~\ref{fig:model-cont}, corresponding to the continuum description, to qualitatively describe the behavior observed in both models. In such scheme, the central dark disc represents the MH particle that, when magnetized, carries a point dipole $\vec \mu_h$ in its center. The orientation of this dipole, corresponding to an arbitrary direction along the magnetic easy axis of the MH particle, defines the symmetry axis of the cell.
The shadowed region around the central disc represents an incompressible elastic shell in which an assembly of implicit MS particles is embedded, whereas the thick solid circle indicates the boundary of this shell when it is unperturbed. Independently of the central particle being magnetized or not, as we showed in Reference~\cite{2019-sanchez-sm2}, when an external field $\vec H_0$ is applied  parallel to the symmetry axis of the system, the elastic shell tends to deform in such a way that its initially circular boundary adopts a prolate shape (dashed line). The prolateness grows with the strength of the field monotonically. When the MH particle is magnetized, a weak external field antiparallel to $\vec \mu_h$ tends to deform the shell into an oblate shape (dotted line). This effect increases with the strength of the field up to a point of maximum oblateness, then decreases until the circular shape is recovered and, finally, an increasing prolateness can be observed again until the critical field required to force the inversion of the central dipole is attained. This latter event is out of the scope of the present study. The scheme in Figure~\ref{fig:model-bs} illustrates how the MH and MS particles are explicitly represented by soft spheres of different size in the bead-spring model, with the same type of arrangement established for the continuum description: a central big sphere representing the MH particle and a spherical shell of small MS particles, all highly interconnected by elastic springs representing the mechanical constraints imposed by the polymer matrix.

In the next Sections we briefly describe each model, underlining the modifications introduced to study the effects of the aforementioned parameters.

\subsection{Continuum analytical approach}
\label{sec:cont_model}
In order to define the analytical equations of our continuum model, here we assume a weak magnetoelastic coupling. Therefore, we can split the magnetoelastic problem into its components, which are described separately in the next two Sections. In addition, the axial symmetry allows us to adopt, without loss of generality, a two-dimensional representation of the system. For both, magnetostatic and elastic contributions, we obtain variational equations that are solved numerically using finite elements. Such calculation is performed with the \emph{FEniCS} computing platform \cite{2015-alnaes-ans}.

\subsubsection{Magnetostatic problem}
We consider the elementary cell outlined in Figure~\ref{fig:model-cont}, placed in an external homogeneous magnetic field $\vec H_0$. As pointed above, the cell consists of a magnetically hard core region, $\Omega_1\ (r<r_1)$, and a magnetically soft shell region, $\Omega_2\ (r_1<r<r_2)$, whereas the empty region external to the cell is denoted as $\Omega_3$. Without loss of generality, $\vec H_0$ points along the $Oz$ axis. The magnetically hard core has magnetization, $M_h$, also coaligned with $Oz$. In this geometry, the magnetically soft shell is a continuum medium that is reversibly magnetizable. Here, we introduce a nonlinear magnetization for this shell by taking the empirical Fr\"{o}lich--Kennelly law \cite{1891-kennelly, 1951-bozorth, 2001-lee-ieeetm},
\begin{equation}\label{eq:FKM}
    \chi(H) = \frac{\chi_0 M^\text{(sat)}}{M^\text{(sat)} + \chi_0 H},
\end{equation}
where $\chi_0$ is the initial susceptibility of the shell and $M^\text{(sat)}$ is its saturation magnetization.

The magnetostatic problem in the absence of charges or currents is described by two Maxwell equations with their respective boundary conditions:
\begin{equation}\label{eq:07}
    \nabla\times{\vec H}=0, \quad \nabla\cdot{\vec B}=0, \quad  [{\vec H}_\tau]=0, \quad [B_n]=0.
\end{equation}
Here, $\vec{B}$ is the magnetic flux density, and subscripts $n$ and $\tau$ denote the components of a vector normal and tangential to the surface of the discontinuity boundary, respectively. Square brackets denote the difference between the values of a quantity on the two sides of the boundary.

The first equation of set (\ref{eq:07}) shows that $\vec{H}$ is a potential field, which can be expressed as a superposition of an external uniform field $\vec{H}_0$ and the gradient of a scalar potential~$\psi$:
\begin{equation}\label{eq:08}
    {\vec H}={\vec H}_0-\nabla\psi.
\end{equation}

With allowance for the rotational symmetry around $Oz$, we may use cylindrical coordinates, so that the potential $\psi$ depends only on the radial distance $\rho$ and axis coordinate $z$. We require that potential $\psi(\rho,z)$, first, vanishes at the external boundary of the cell: $\psi \big|_{\rho=r_2}=0$; and second, is periodic along the $Oz$ axis: $ \psi(\rho, z)= \psi(\rho,\,z+h)$, where $h$ is the cell period.

The solution of equations~(\ref{eq:07}) is equivalent to finding an extremum of the energy functional for the entire volume $V$ \cite{1963-landau-lifshitz}:
\begin{equation}\label{eq:09}
\int_V {\vec{B}}\cdot\delta {\vec{H}} \,dV=0,
\end{equation}
where here and hereinafter $\delta$ denotes the variation of a quantity.

The vector of magnetic flux density is defined in each region of the model cell as follows:
\begin{equation}\label{eq:10}
    \vec{B} = \begin{cases}
      {\vec{H}}_0 - \nabla{\psi} + 4 \pi{\vec{M}}_h &\text{ for } \Omega_1, \\
      \left(1+4\pi\chi({H})\right)\left({\vec{H}}_0-\nabla{\psi}\right) &\text{ for }  \Omega_2,\\
      {\vec{H}}_0 - \nabla{\psi} &\text{ for }  \Omega_3.
    \end{cases}
\end{equation}

Taking into account the relations:
\begin{equation*}
 \delta{\vec{H}}=-\nabla\delta{\psi} \quad \text{and} \quad \int_V {\vec{H}}_0\cdot\nabla\delta{\psi}\,dV=0,
\end{equation*} 
equation (\ref{eq:09}) can be transformed into
\begin{multline}\label{eq:11}
\frac1{4\pi}\!\int_{V}\!\nabla{\psi}\cdot\nabla\delta{\psi}\,dV\!=\!\int_{V_{\Omega_2}}\!\!{{\vec{M}}}_h\!\cdot\!\nabla\delta{\psi}\,dV \!+ \\
\!\int_{V_{\Omega_3}}\!\!\chi(H)\left({\vec{H}}_0\!-\!\nabla{\psi}\right)\!\cdot\!\nabla\delta{\psi}\,dV.
\end{multline}
Finally, this variational equation is solved numerically for the potential $\psi(\rho,z)$, finding solutions for each sampled value of the initial susceptibility, $\chi_0$.

\subsubsection{Elastic problem}
Having once obtained the solution of the magnetostatic problem, {\it i.e.}, having found the distribution of the magnetic field inside the magnetically soft shell, one can calculate how the shell would deform under the resulting magnetic forces. In order to do that, we need to formulate the equations for a magnetoelastic medium, assuming balance between magnetic and elastic forces:
\begin{equation}
   \nabla \cdot \tens \sigma + \nabla \cdot \tens \sigma_\text{m} = 0,
   \label{eq:mag-el}
\end{equation}
where $\tens\sigma$ denotes the elastic stress tensor and $\tens\sigma_\text{m} = \frac1{4\pi} \vec B \vec H - \frac1{8\pi} H^2 \tens g$ is the Maxwell stress tensor. In case of equilibrium, the pressure on both sides of the outer boundary $\Gamma$ should be the same. Thus, one obtains:
\begin{equation}
   \vec n \cdot \tens \sigma \big|_{\Gamma} 
   = \vec n \cdot(\tens\sigma_\text{m}^\text{(e)} - \tens\sigma_\text{m}^\text{(i)})\big|_{\Gamma}
    = 2\pi M_n^2\vec n \big|_{\Gamma},
\label{eq:pres}
\end{equation}
where $\vec n$ is the vector normal to the outer boundary and $(.)^\text{(i)}$ and $(.)^\text{(e)}$ denote internal (inside the shell) and external (outside the shell) values. Then we write Hooke's law and the relation between strain tensor $\tens e$ and displacement vector $\vec u$ as
\begin{equation}
   \tens\sigma = \lambda {\rm tr}(\tens e)\tens g + 2G\tens e, \quad
   \tens e = \frac12 (\nabla\vec u + \nabla\vec u^{\rm T}),
\end{equation}
where $\tens g$ is unity tensor, $G$ stands for the shear modulus, and the Lam{\'e} coefficient $\lambda$ characterizes the compressibility of the material, and is related to its volume elastic modulus as $K = \lambda + 2G/3$.

In order to obtain a variational form of the magnetoelastic problem using the principle of virtual work, we have to multiply equations \eqref{eq:mag-el} and \eqref{eq:pres} by $\delta\vec u$ and integrate: 
\begin{multline}
    \int_V \left(\nabla\cdot\tens\sigma + \nabla\cdot\tens\sigma_\text{m}^\text{(i)}\right)\cdot\delta\vec u dV - \\
    \int_S \vec n \cdot \left(\tens\sigma + \tens\sigma_\text{m}^\text{(i)} - \tens\sigma_\text{m}^\text{(e)}\right) \cdot\delta\vec u dS = 0.
\end{multline}

Employing Gauss-Ostrogradsky theorem, after simplifications, we come to a so-called weak variational form:
\begin{multline}
    \int_V \left(
        \lambda {\rm tr}(\tens e) {\rm tr}(\delta\tens e) + 2 G\tens e\cdot\cdot\delta\tens e +  \tens\sigma_\text{m}^\text{(i)}\cdot\cdot\delta\tens e
        \right) dV = \\
        = \int_S (2\pi M_n^2\vec n + \vec n \cdot \tens\sigma_\text{m}^\text{(i)})\cdot\delta\vec u\, dS.
\end{multline}

As pointed above, the presence of a magnetic field transforms our initially spherically symmetric problem into an axisymmetic one. Therefore, here we also use cylindrical coordinates, $(\rho,\,z)$, and solve the problem numerically with finite elements, obtaining $\vec u (\rho, z)$ for the quarter of the main cross-section of the cell. In order to do this, we need to apply the Dirichlet boundary conditions
\begin{equation}
    u_\rho\big|_{\rho=0} = 0, \quad u_z\big|_{z=0} = 0,  \quad \vec u|_{r=r_1} = 0,
\end{equation}
which mean that the shell is immobile at the shell-core boundary, and the symmetry requirement applies at the boundaries $\rho=0$ and $z=0$. For all calculations, the ratio $\lambda/G = 10^3$ was fixed.

\subsection{Bead-spring model}
Our bead-spring model is designed for molecular dynamics (MD) simulations. The solid magnetic particles are represented as soft spheres, assuming that they are always surrounded by an elastic layer of polymer material that prevents them to get into close contact. This assumption is consistent with the weak magnetoelastic coupling established for the continuum model. The soft core pair interaction is defined by a truncated and shifted Lennard-Jones potential, aslo known as Weeks-Chandler-Andersen (WCA) interaction \cite{1971-weeks}:
\begin{equation}
U_{\mathrm{{WCA}}}(r)= \left\{ \begin{array}{ll}
U_{\mathrm{LJ}}(r)-U_{\mathrm{LJ}}(r=r_{\mathrm{cut}}), & r<r_{\mathrm{{cut}}}\\
0, & r\geq r_{\mathrm{{cut}}}
\end{array}\right. ,
\label{eq:WCA}
\end{equation}w
where  $r=\| \vec r_i - \vec r_j \|$ is the center-to-center distance between the pair of particles $i$ and $j$, $U(r) = 4 \epsilon_{\mathrm{LJ}} \left [ (d / r )^{12} - (d/r)^6\right ]$ is the conventional Lennard-Jones potential, $r_c$ is the truncation length, set to $r_c=2^{1/6} d$ in order to make the interaction purely repulsive, and $d$ is the center-to-center excluded distance, defined by the characteristic diameter of each particle, $d_i$ and $d_j$, as $d=(d_i + d_j)/2$.

As mentioned above, the mechanical constraints imposed by the polymer matrix are represented by a network of elastic springs, with a simple harmonic potential
\begin{equation}
U_{\mathrm{S},i}(r)=\tfrac12{k_i}(r-L_i)^2,
\label{eq:u-spring}
\end{equation}
where $r$ is the distance between the connected points, $k_i$ is the elastic constant of the spring and $L_i$ its equilibrium length. The connections points are the centers of the MS particles and a set of fixed anchoring points randomly distributed on the surface of the MH particle, that remains permanently immobile. In order to ease the fitting of the elastic properties of this spring network, we take the elastic constants to be proportional their corresponding equilibrium lengths, with the average equilibrium length of all the springs, $\langle L \rangle$, as scaling factor:
\begin{equation}
 k_i = \bar k \frac{L_i}{\langle L \rangle}.
 \label{eq:elast-const}
\end{equation}
In this way, the only fitting parameter for the whole network is the constant factor $\bar k$. The direct comparison of the deformations obtained in the continuum and the bead-spring model \cite{2019-sanchez-sm2} showed that the spring network of the latter fits rather well the elastic properties defined for simple mass-spring (MS) networks \cite{2015-kot}. The bulk modulus of such networks is
\begin{equation}
K_{\textsc{ms}}=\frac{n \langle S \rangle \langle k L^2 \rangle}{18}=\frac{n  \bar k \langle S \rangle \langle L^3 \rangle}{18 \langle L \rangle},
\label{eq:bulk_kot}
\end{equation}
where $n$ is the number density of connection points and $\langle S \rangle$ is the average number of springs connected each of them. Assuming spatial isotropy and a Poisson ratio for the simple mass-spring network of $\nu=1/4$, then the shear modulus can be defined as
\begin{equation}
G_{\textsc{ms}}\!=\!\frac{3K_{\textsc{sb}}(1-2\nu)}{2(1+\nu)}\!=\!\frac{n \langle S \rangle \langle k L^2 \rangle}{30}=\frac{n \bar k \langle S \rangle \langle L^3 \rangle}{30 \langle L \rangle}.
\label{eq:mod_kot}
\end{equation}

The magnetic properties of the particles are represented as point dipoles located at their centers. The moment of the dipole corresponding to the MH particle, $\vec \mu_h = \mu_h \hat k$, is fixed to $\mu_h=M_h V_h$, where $M_h$ is its magnetization, that we take as constant, and $V_h$ its volume. In the same way, the dipole moment of the $i$-th MS particle is given by
\begin{equation}
 \vec \mu_i = \Mfk V_s,
\end{equation}
where $V_s$ is its volume and $\Mfk$ its magnetization. In this case, according to its magnetically soft nature, $\Mfk$ is defined as 
\begin{equation}
 \Mfk= \chi_i \Hint,
\end{equation}
where $\chi_i$ is the field dependent susceptibility and $\Hint$ the internal field inside the particle, which is parallel to the net external field at its position, $\Hext$. Following the Fr\"{o}lich--Kennelly nonlinear magnetization introduced above for the continuum model, $\chi_i$ is given by equation~(\ref{eq:FKM}) and the modulus of the internal field, $\mHint=\| \Hint\| $, is given by $\mHext = \| \Hext \|$ as
\begin{equation}
 \mHint = \mHext - \frac{4 \pi}{3} \frac{\Xm \mHint }{1 + \frac{\Xm}{\Msat} \mHint},
 \label{eq:FKintfield}
\end{equation}
where $\Xm$ is the initial susceptibility of the MS material and $\Msat$ its saturation magnetization. From this expression we obtain:
\begin{multline}
    \mHint = \frac{1}{6 \Xm} \left [ 3  \Xm \mHext - 4 \pi  \Xm \Msat - 3 \Msat + \right . \\ + \left ( 9 \Xm^2 \mHext^2 - 24 \pi \Xm^2 \mHext \Msat + 18 \Xm \mHext \Msat + \right . \\ \left . \left . + 16 \pi^2 \Xm^2 \Msat^2 + 24 \pi \Xm \Msat^2 + 9 \Msat^2 \right )^{1/2} \right ].
\end{multline}
Here we consider only two contributions to $\Hext$: the externally applied field, $\vec H_0$, and the field generated by the dipole of the MH particle, when it is magnetized, at the center of the MS one, $\vec H_h^{(i)}$,
\begin{equation}
 \Hext = \vec H_{0} + \vec H_h^{(i)}.
\end{equation}
The latter is defined as
\begin{equation}
 \vec H_h^{(i)}=\frac{3 \vec r_i (\vec \mu_{h} \cdot \vec r_i)}{r_i^5} - \frac{\vec \mu_h}{r_i^3},
\end{equation}
where $\vec r_i$ is the vector connecting the center of the MH particle to the center of the polarized one and $r_i=\| \vec r_i\|$. In this way, we disregard mutual magnetization between MS particles when calculating their induced dipoles. However, we fully take into account the dipole-dipole interaction between any pair of magnetized particles,
\begin{equation}\label{eq:dip-dip}
U_{dd}(i j) =-   3 \frac{ \left ( {\vec \mu_i }
\cdot {\vec r_{ij}} \right ) \left ( {\vec \mu_j }\cdot {\vec
r_{ij}} \right ) }{r^5} + \frac{ \left ( {\vec \mu_i } \cdot {\vec \mu_j } \right ) }{r^3},
\end{equation}
where $\mu_i$, $\mu_j$ are their respective dipole moments, ${\vec r_{ij}} = {\vec r_i} - {\vec r_j}$ is the vector connecting their centers and $r=\| \vec r_{ij} \|$. Finally, MS particles also experience the Zeeman interaction with the external applied field. However, since their dipoles are induced, the effective interaction corresponds to one half of the conventional Zeeman potential \cite{1963-landau-lifshitz}:
\begin{equation}
 U_{H} = - \frac{1}{2} \vec \mu_i \cdot \vec H_{0}.
\end{equation}

We perform our simulations with the package {ESPResSo} 4.1 \cite{2019-weik-epjst}, using MD with a Langevin thermostat \cite{2007-berendsen}. Therefore, we perform Langevin dynamics (LD) simulations, integrating the Langevin translational and rotational equations of motion with the Velocity Verlet algorithm \cite{2004-rapaport, 2007-berendsen}. In difference with most usual LD simulations, we work under a quasi-athermal regime by setting a very small thermal energy in the system---around $10^2$ times smaller than the average elastic energy of each spring under deformations produced by moderate applied fields. Therefore, our simulations correspond to an energy minimization with slight thermal fluctuations. The latter help the system to relax without getting trapped into high energy local minima.

As is usual in coarse-grained simulations, we use a system of dimensionless units. Hereinafter, we denote dimensionless parameters with a tilde symbol, $\tilde X$. We take the diameter of the MS particles as reference length scale, so that $\tilde d_s=1$ and $\tilde d_h=10$, and the shear modulus of the matrix, $G$, as the reference scale for magnetic parameters, so that dimensionless field is defined as $\tilde H = H / \sqrt{G}$, magnetization as $\tilde M = M / \sqrt{G}$ and dipole moment as $\tilde \mu = \mu / \sqrt{G} d_s^3$. The latter definitions also apply to the results of the continuum model. Since here we are not interested in dynamics, for simplicity we take the Langevin translational and rotational friction coefficients as unity, and the thermal energy.

Each simulation run starts by placing and fixing the position and orientation of the MH particle in a simulation box with open boundaries. $N_a=99$ connection points for the springs are randomly assigned to its surface. Around it, $N_s=2 \cdot 10^3$ MS particles are randomly placed inside a spherical shell of dimensionless thickness 5. The latter are let to relax inside the shell by simply following their steric interactions, so any soft core overlap is removed. Then the spring network is build up by randomly choosing pairs of connecting points according to the following rules: first, the distance between them is not larger than $d_{\mathrm{cut}}=6$; second, none of them has more than $s_{\mathrm{max}}=6$ springs attached. These arbitrary rules provide a good compromise between locality and isotropy of the elastic constraints acting of each MS particle, in one hand, and the computational load, on the other. The result of such build up procedure is a highly connected network with $\langle S \rangle \approx 6$, $\langle \tilde L \rangle \approx 4$ and $\langle \tilde L^3 \rangle \approx 93$ \cite{2019-sanchez-sm2}. Taking into account that the dimensionless number density of connecting points is $\tilde n = 6 (N_s + N_a) / 7 \pi {\tilde d}_h^3$, we can use equation (\ref{eq:mod_kot}) in dimensionless units to fit the elastic prefactor $\bar k$:
\begin{equation}
 \bar k = \frac{30 \tilde G \langle \tilde L \rangle}{\tilde n \langle S \rangle \langle \tilde L^3\rangle} = \frac{35 \pi \tilde G \tilde d_h^3 \langle \tilde L \rangle}{(N_s+N_a) \langle S \rangle \langle \tilde L^3\rangle } \approx 0.4.
\end{equation}

Finally, with the setup described above, we set the central dipole $\tilde \mu_h$ and external field $\tilde H_0$ to their selected values, performing a final relaxation run of $5 \cdot 10^5$ integration steps, using a timestep $\delta \tilde t = 0.01$. Only the final configuration obtained from each run is analyzed. For each set of sampled parameters, statistics are collected from 60 runs with independent initial configurations. In this case, we only sample different fields for two cases: non magnetized central particle, $\tilde \mu_h=0$, and magnetized central particle with the lowest sampled matrix rigidity, $G=10^5$~dyn/cm$^3$, corresponding to $\tilde \mu_h=1324.6$.

\section{Results and discussion}
We start the discussion by considering the simplest case, that is when the central particle in the HME elementary unit is nonmagnetic: $\tilde \mu_h=0$.
The first task is to find a correct common basis to compare the magnetically induced deformations predicted by the continuum and the bead-spring models.
Whereas in the former the outer edge of the matrix is perfectly defined and the deformations are easy to visualize, in the bead-spring model no explicit outer boundary exists (see Figure~\ref{fig:model-bs}) since it is rendered by the discrete positions of MS particles.
To find commensurate terms for that comparison, we define a virtual boundary of the bead-spring system as follows.
First, the convex hull of all particles in the system is calculated. 
Then, by assuming that under any moderate deformation the elastic shell keeps an ellipsoidal profile, we perform a least-squares fit of an ellipsoid to that convex hull.

Taking advantage of the afore-introduced ``ellipsoid terms'', we characterize the deformations of the shell boundary by means of a single parameter, defined as $\Delta c^* = \left \langle  (c - c_0)/c_0 \right \rangle$, where $c$ is the distance (along the line parallel to the external field) from the center of the MH particle to the point where this line intersects the outer shell boundary , $c_0$ is the value of that distance when $\tilde \mu_h=0$ and $\tilde H_0=0$; angle brackets denote the average over independent runs.
Thus, $\Delta c^*$ is positive for stretching of the cell along the field and negative in opposite case.

\begin{figure}[!h]
\centering
\includegraphics[width=8.cm]{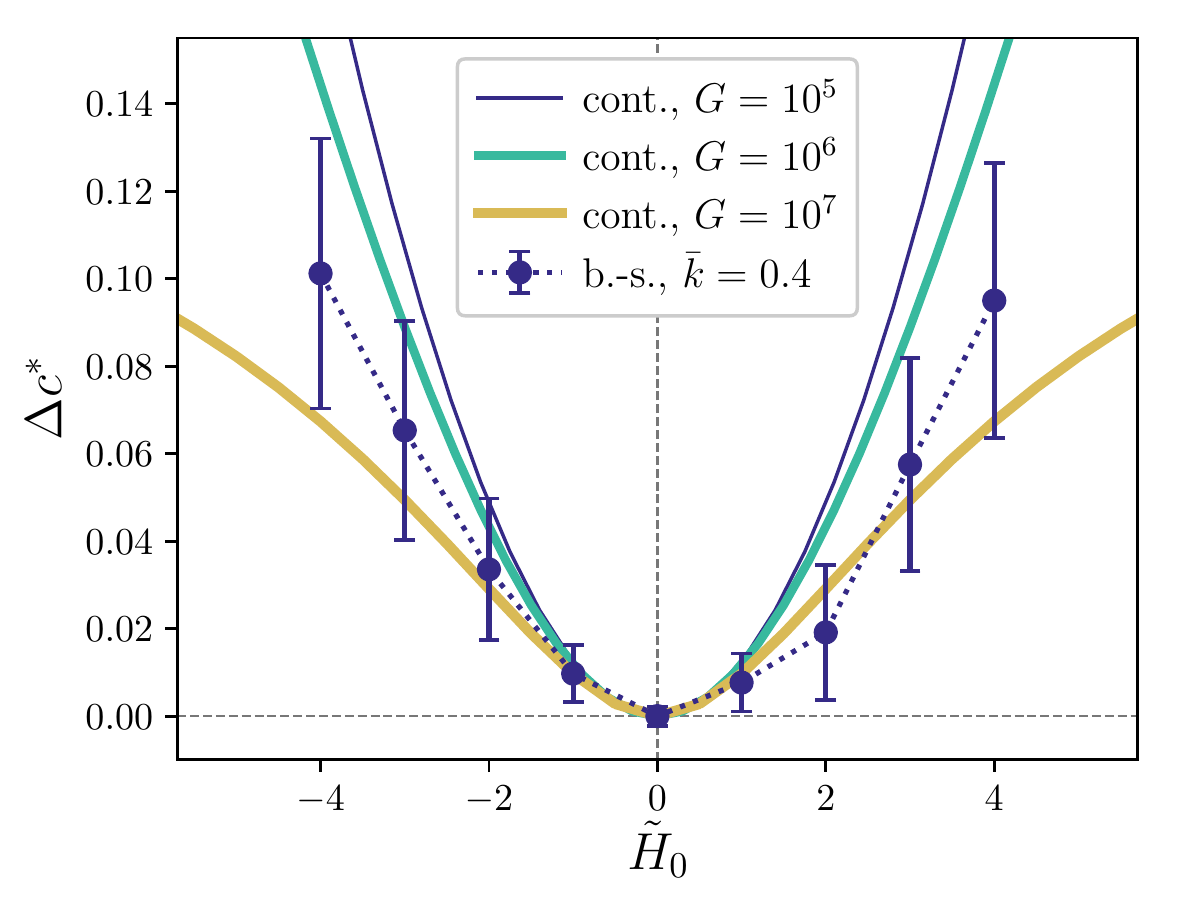}
\caption{Longitudinal deformation parameter, $\Delta c^*$, as a function of the applied field, $\tilde H_0$, for the case of nonmagnetized central particle, $\tilde \mu_h=0$. Results of the continuum model for different values of the elastic modulus $G$ are rendered by solid lines, data provided by the bead-spring model for $\bar k =0.4$ are shown by symbols with error bars. Dotted curve connecting the symbols is a guide for the eye.}
\label{fig:Gs-dmc0}
\end{figure}

Figure~\ref{fig:Gs-dmc0} shows the dependence of $\Delta c^*$ on $\tilde H_0$ for both models at $\tilde \mu_h = 0$.
Note that the sign of $\tilde H_0$ indicates its orientation with respect to the reference axis.
The curves are perfectly symmetric with respect to the ordinate axis ($\tilde H_0=0$), \textit{i.e.}, the unperturbed state of the system.

The results of the continuum model presented in Figure~\ref{fig:Gs-dmc0} correspond to several values of the elasticity modulus $G$.
Note that the curves are plotted vs nondimensional field magnitude $\tilde H_0 = H_0/\sqrt{G}$, so that one and the same abscissa point at different $G$'s renders different dimensional values of the field.
Had the calculation been done with the linear magnetization law ($M^\text{(sat)} = \infty$ in Eq.~(\ref{eq:FKintfield})), all the curves would have coincided.
However, the nonlinear magnetization dependence (that we allow for here) removes this degeneration since the saturation magnetization is scaled with $\sqrt{G}$ as well.
Due to that, the nondimensional saturation magnetization is lower for stiffer matrices and, as a result, the nonlinearities become more distinct.
This is easily visible in Figure~\ref{fig:Gs-dmc0}, where the curves are presented, which have been calculated for $\tilde M^\text{(sat)} = \frac{1500\, \text{emu}/cm^3}{\sqrt G}$, see the curve for $G=10^{7}$\,dyn/cm$^3$.
The curve rendered by the bead spring-model has been calculated for parameter $\tilde k=0.4$, and the comparison implies that in the considered case of $\tilde \mu_h = 0$, the effective modulus that one may attribute to this system should lie inside the interval $10^6-10^7$~dyn/cm$^3$.

\begin{figure}[!h]
\centering
\includegraphics[width=8.75cm]{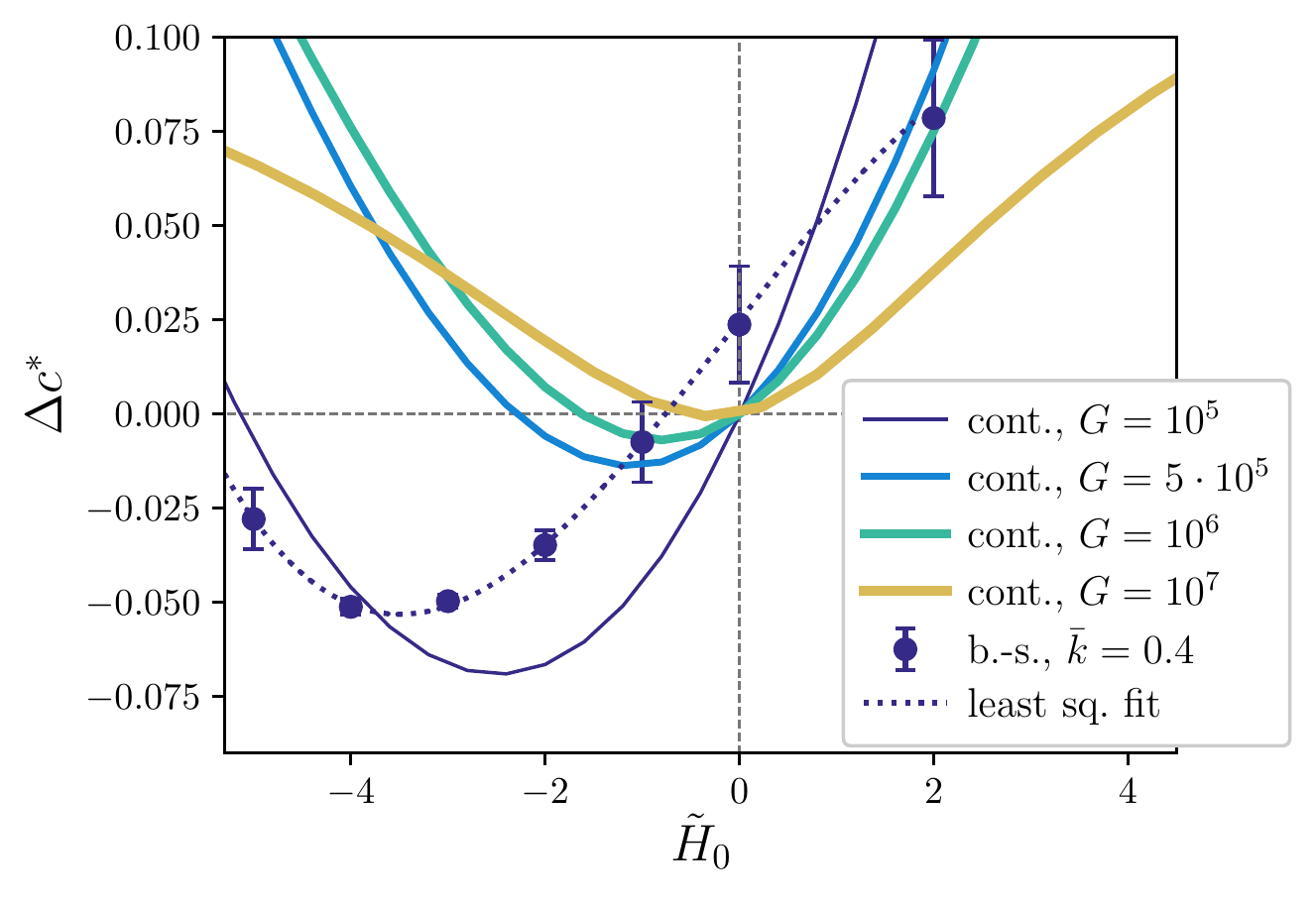}
\caption{Longitudinal deformation parameter, $\Delta c^*$, as a function of the applied field, $\tilde H_0$, when the central particle has a magnetization of 800 emu/cm$^3$. Solid lines correspond to the results of the continuum model for different values of the elastic modulus, $G$, symbols with error bars to the results of the bead-spring model for $\bar k =0.4$. Dotted curve is a cubic splines fit to the latter.}
\label{fig:Gs-dmc1}
\end{figure}

Figure~\ref{fig:Gs-dmc1} shows the results on $\Delta c^* (\tilde H_0)$ obtained when the central particle in the system bears permanent magnetic moment $\tilde \mu_h = \frac{4\pi\, 800}{3\sqrt G}$, where $4\pi/3$ is the nondimensional volume of the MH particle and $800$\,emu/cm$^3$ its magnetization.
In this case the results of both models, although qualitatively similar, are quantitatively rather different.
In both approaches, the essential effect of the magnetic field of the MH particle is to shift the minimum of the parabolic profile to negative values, thus producing oblate shape of the cell under inverted field.
As already explained, with the employed scaling scheme, the increase of elastic modulus entails reduction of all the magnetic contribution, and this no surprise that this shift becomes smaller.
This tendency is clearly visible if to compare with one another the curves rendered by continuum model in Figure~\ref{fig:Gs-dmc0}.

It is instructive, however, to compare the continuum model with the bead-spring one.
The latter demonstrates an ``ambivalent'' behavior, as follows from Figure~\ref{fig:Gs-dmc1}.
Indeed, under negative field the bead-spring model curve resembles that of the continuum one with elastic modulus about $10^5$\,dyn/cm$^3$ whereas under positive magnetization it rather displays similarity with the continuum curves corresponding to much higher elasticities: $G\sim10^6-10^7$\,dyn/cm$^3$.

To understand such field-tuned softening/stiffening, we recall that at $H_0<0$ the external field around the poles of the MH particle substantially compensates the field of the core and, thus, the field in the shell is on the average reduced.
Under those conditions, the MS particles are less magnetized that entails lower aggregation and, consequently, makes the bead-spring shell becomes effectively more soft.
When the external field is in the $H>0$ range, in the ``polar'' zones the core field adds to the external one.
This makes the MS particle aggregation in those zones stronger that, in turn, induces higher stresses inside the inter-bead spring mech and, by that, reduces its ability to deformations.
As a consequence, the overall stiffness of the shell increases, as it is seen from comparison of the bead-spring (dashed) and continuum (solid) curves in the right-hand part of Figure~\ref{fig:Gs-dmc1}.
The field-modulated elastic modulus is an essential feature of the bead-spring model; note that in the continuum consideration such an effect is entirely absent.
Meanwhile, as Figure~\ref{fig:Gs-dmc1} shows, the contribution of field-tuning -- and this effect is most probably present in real magnetic elastomers -- turns out to be sufficiently strong, and because of that appeals for further investigation.

\begin{figure}[!h]
\centering
\includegraphics[width=8.cm]{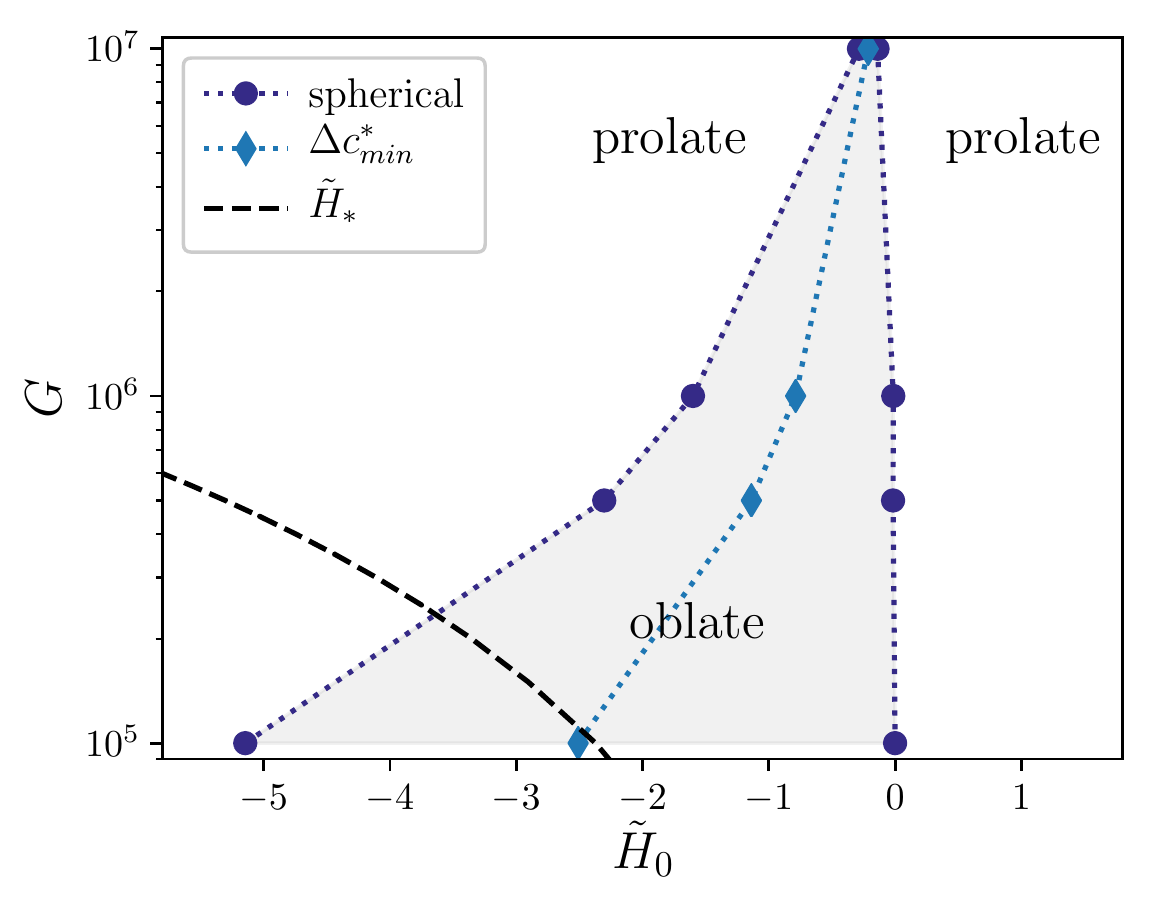}
\caption{Deformation diagram of the \HME\ elementary unit as a function of the applied field, $\tilde H _0$, and the elastic modulus of the matrix, $G$, as predicted by the continuum model. Shadowed region corresponds to oblate deformations, being delimited by curves of no effective deformation (filled circles). Curve of filled diamonds indicates the maximum oblate deformation. Outside the oblate region, the system deforms into prolate shapes. Dashed curve is the critical field $\tilde H_*$ that would invert the orientation of the central dipole moment: results on the left of this curve are unphysical.}
\label{fig:reentrant-chi025}
\end{figure}

Figure~\ref{fig:reentrant-chi025} presents the dependence of the field-induced anisometricity of the considered cell on the, this time dimensional, values of elasticity modulus; here only the results of continuum model are presented.
At this diagram, the shaded curvilinear triangle corresponds to the combination of parameters under which the cell is oblate; in the points that make the borders of the triangle, the cell is spherical; outside the shaded area the cell responds to the  applied field by elongation.
As expected, with the increase of elastic modulus, the region of oblateness becomes more narrow and virtually disappears at about $G=10^7$\,dyn/cm$^3$.
Note also that the dominating part of the triangle lies to the right of the dashed line that corresponds to the magnetic switching of the MH core of the shell.

When relating the above-presented results to a real situation, one essential issue is to be clarified concerning the response of the MH core to the inverted (negative, in our notation) field.
Indeed, under such a field a particle with permanent magnetic moment $\vec{\mu}_h$ residing in a compliant matrix is well able to rotate mechanically in order to turn $\vec{\mu}_h$ to the actual direction of the field.
To get an estimation for the negative field strength when it happens, we consider a spherical single-domain particle of uniform magnetization $M_h$ sitting in an elastic environment of shear modulus $G$.
For the orientational-dependent energy of the system in the inverse field (antiparallel to ${\vec\mu}_h$) we have
\begin{equation}
U=M_hHV_p\cos\vartheta+3GV_p\vartheta^2,
\label{eq:Uorient}
\end{equation}
where 
$V_p$ is the particle volume.
As the particle is single-domain and magnetically hard, the magnetic moment is ``frozen'' into its body, so that the angle $\vartheta$ describes simultaneously: the deviation of the magnetic moment from the direction of the field and the angular displacement of the particle from its initial position.

Expanding (\ref{eq:Uorient}) for small angular perturbations ($\vartheta=0$ is the initial state), one gets
\begin{equation*}
U\approx M_hHV_p\left(1-{\textstyle\frac12}\vartheta^2+{\textstyle\frac{1}{24}}\vartheta^4\right)+3GV_p\vartheta^2.
\end{equation*}
Differentiation respect to $\vartheta$ yields
\begin{equation*}
\partial U/\partial\vartheta \approx-M_hH\vartheta\left[1-{\textstyle\frac16}\vartheta^2\right]+6G\vartheta,
\end{equation*}
so that the condition of minimum writes
\begin{equation}\label{eq:thetamin}
\vartheta\left[6G-M_hH\left(1-{\textstyle\frac16}\vartheta^2\right)\right]=0.
\end{equation}

From (\ref{eq:thetamin}) it is easy to find out that the particle dwells in the initial state $\vartheta=0$ at $H<H_\ast$ and acquires a nonzero angle deviation (commences mechanical rotation) at $H>H_\ast$ with the critical field $H_\ast=6G/M_h$; in nondimensional form it is ${\tilde H}_\ast=6\sqrt{G}/{\tilde M}_h$.
The latter dependence is plotted in Figure \ref{fig:reentrant-chi025} by dashed line.
According to the definition of $H>H_\ast$, the region $|{\tilde H}|>{\tilde H}_\ast$ (to the left) is unphysical since there the particle mechanical rotation should occur.
At $M_h=800$\,emu at the lowest used here value $G=10^5$\,dyn/cm$^2$ the critical field is ${\tilde H}_\ast(G)=-2.37$, see the point on the abscissa axis of Figure \ref{fig:reentrant-chi025}.
As seen, the minima of the presented curves and the full re-entrant shrinking effect are justified only for the cells with elasticity $G>6\cdot10^5$\,dyn/cm$^2$.
Although the existence of the minima (see Figure \ref{fig:Gs-dmc1}) of the bead-spring model and the continuum one for $G=10^5$\,dyn/cm$^2$ doubtful, the occurrence of negative cell shrinking effect falls well in the physical region.

It is worth of noting, however, that afore-obtained expression, in fact, underestimates the strength of the inverse field if to refer to the external applied one.
Indeed, according to its derivation, ${\tilde H}_\ast$ is the field experienced by the MH core of the cell.
Due to that, the absolute value of minimal external field strength capable of initiating the particle rotation, exceeds ${\tilde H}_\ast$ by the strength of the demagnetizing field generated in the MS shell.
This means that if to transform the scaling of abscissa in Figure \ref{fig:reentrant-chi025} to the units of external magnetic field, the dashed line would shift yet further to the left, thus yet widening the range of applicability of our model.

\section*{acknowledgements}
P.A.S. and S.S.K. acknowledge support by the DFG grant OD 18/24-1, by the Act 211 of the Government of the Russian Federation, contract No.~02.A03.21.0006, and by the FWF START-Projekt Y 627-N27. O.V.S. and Yu.L.R. acknowledge support by RFBR project 19-52-12045. Computer simulations were carried out at the Vienna Scientific Cluster.


\end{document}